\documentclass[fleqn,12pt]{wlscirep}

\usepackage{color,soul}
\usepackage{upgreek}
\usepackage{mathastext}
\usepackage{setspace}
\usepackage{eqnarray,amsmath}                     

\title{\centering  A single metagrating metastructure for wave-based parallel analog computing}

\author[1]{ Hamid Rajabalipanah}
\author[2]{Ali Momeni}
\author[3]{Mahdi Rahmanzadeh}
\author[1]{Ali Abdolali}
\author[2]{Romain Fleury}
\affil[1]{Applied Electromagnetic Laboratory, School of Electrical Engineering, Iran University of Science and Technology, Tehran 1684613114, Iran.}
\affil[2]{Laboratory of Wave Engineering, School of Electrical Engineering,
Swiss Federal Institute of Technology in Lausanne (EPFL), Lausanne, Switzerland.}
\affil[3]{Electrical Engineering Department, Sharif University of Technology, Tehran 11155-4363, Iran.}

\doublespace 

\begin{abstract}
Wave-based signal processing has witnessed a significant expansion of interest in a variety of science and engineering disciplines, as it provides new opportunities for achieving high-speed and low-power operations. Although flat optics desires integrable components to perform multiple missions, yet, the current wave-based analog computers can engineer only the spatial content of the input signal where the processed signal obeys the traditional version of Snell's law. In this paper, we propose a multi-functional metagrating to modulate both spatial and angular properties of the input signal whereby both symmetric and asymmetric optical transfer functions are realized using high-order space harmonics. The performance of the designed compound metallic grating is validated through several investigations where closed-form expressions are suggested to extract the phase and amplitude information of the diffractive modes. Several illustrative examples are demonstrated to show that the proposed metagrating allows for simultaneous parallel analog computing tasks such as first- and second-order spatial differentiation through a single multi-channel structured surface. It is anticipated that the designed platform brings a new twist to the field of optical signal processing and opens up large perspectives for simple integrated image processing systems. 

\end{abstract}
\begin{document}
\flushbottom
\maketitle
%
%
\thispagestyle{empty}
\section{Introduction}

For a few decades, digital processors have been widely used to execute computational tasks, as an alternative to analog mechanical and electrical computers. Despite their reliability and high-speed operation, digital processors suffer from high-power consumption, expensive
analog-to-digital conversion, and sharp performance degradation at high frequencies, leading to large limitations even for performing simple computing tasks such as differentiation or integration, equation solving, matrix inversion, edge detection, and image processing \cite{zangeneh2020analogue,abdollahramezani2020meta}. With the advent of metamaterials and metasurfaces, spatial analog optical computing resurfaced, finding important applications as compact solutions for high speed, high throughput image processing and parallel computing. Since the seminal proposal of Silva \textit{et al.,\cite{silva2014performing}} wave-based analog computing has witnessed rapid progress, with the demonstration of optical spatial differentiators \cite{zhou2020flat,zhu2017plasmonic,zhou2019optical,sol2021meta,momeni2019generalized,zhou2020analog, momeni2020reciprocal,momeni2021asymmetric,kwon2018nonlocal,davis2019metasurfaces}, integrators \cite{abdollahramezani2020meta,rajabalipanah2021analog,momeni2019generalized, kwon2018nonlocal}, equation solvers \cite{estakhri2019inverse,camacho2021single,abdolali2019parallel}, spatiotemporal computing \cite{zhou2021analogue,momeni2021switchable} and wave-based neuromorphic computing \cite{lin2018all,hughes2019wave,zuo2019all,momeni2021wave} . Among them, the Green’s function (GF) method, in which a specific-purpose computing operation is directly realized in real space, without transforming back and forth from the spatial to the spectral domain, affords compactness and avoids possible challenges in error propagation and alignment issues. The applicability of the GF method to execute signal processing has been verified in a series of proposals via spin hall effect of light \cite{zhu2019generalized}, disordered and complex scattering system \cite{sol2021meta,matthes2019optical,del2018leveraging}, layered structures \cite{zhou2021analogue,jin2021transmissive}, topological insulators \cite{zangeneh2019topological,zangeneh2020disorder}, plasmonic arrays \cite{zhu2017plasmonic}, bianisotropic metasurfaces \cite{momeni2020reciprocal,abdolali2019parallel}, and so on. Nevertheless, prior GF-based studies still face two different challenges: (i) parallel realization of mathematical operators has been only addressed by using bulky structures \cite{camacho2021single} and array of subwavelength meta-atoms with complex geometries \cite{ babaee2021parallel,babaee2020parallel} and thus, they are still subject to implementation difficulties arising from high fabrication precision demands; (ii) although reflective optical processing for normal incidences is a good alternative for complex oblique illumination setups, it still needs additional optical components to separate the processed signal from the input one \cite{momeni2020reciprocal}. Further efforts to tackle these barriers must be accompanied with the use of more powerful architectures to implement spatial optical signal processing.

Passive metasurfaces can ensure highly-efficient wavefront molding by leveraging nonlocal effects stemming from the excitation of evanescent \cite{wang2018extreme,li2020harnessing} or leaky modes \cite{abdo2018leaky}, but this typically leads to complex design requirements and a need for deeply subwavelength fabrication resolution \cite{yang2014all,yu2015high,kamali2018review,kiani2020self,kiani2020spatial,rajabalipanah2019asymmetric,rajabalipanah2019addition,momeni2018information,rajabalipanah2020real,rouhi2019multi}. As an alternative solution, recent efforts have shown that metagratings composed of non-subwavelength periodic patterns enable highly complex diffraction scenarios of significant practical interest, with high efficiency \cite{ra2017metagratings,ra2018reconfigurable}. The core idea of metagratings is based on Floquet-Bloch (FB) theory, remarking that when a plane wave impinges on a periodic
structure, a discrete set of diffracted waves can be generated,  some of them propagating and others being evanescent \cite{rahmanzadeh2020perfect,rahmanzadeh2021analytical}. The number of propagating and evanescent waves is determined by the period of the structure and the angle of incidence. The structure is engineered at the scale of the wavelength so as to suppress the unwanted space harmonics and reroute the incident power towards a desired non-specular channel. Therefore, such design potentially relaxes some of the fabrication challenge of metasurfaces, as it does not require precise lithography techniques \cite{ra2017metagratings,sell2017periodic}. Metagratings have also the ability to provide multiple arbitrarily-oriented space channels for creating multi-mission surfaces \cite{sell2017large}. Although several reports have examined metagratings from different points of view, the potential application of these structures for performing optical analog computation, which requires studying the phase information of the space harmonics, has not been unveiled, yet.   
\begin{figure}[h!]
	\centering
	\includegraphics[width=\textwidth]{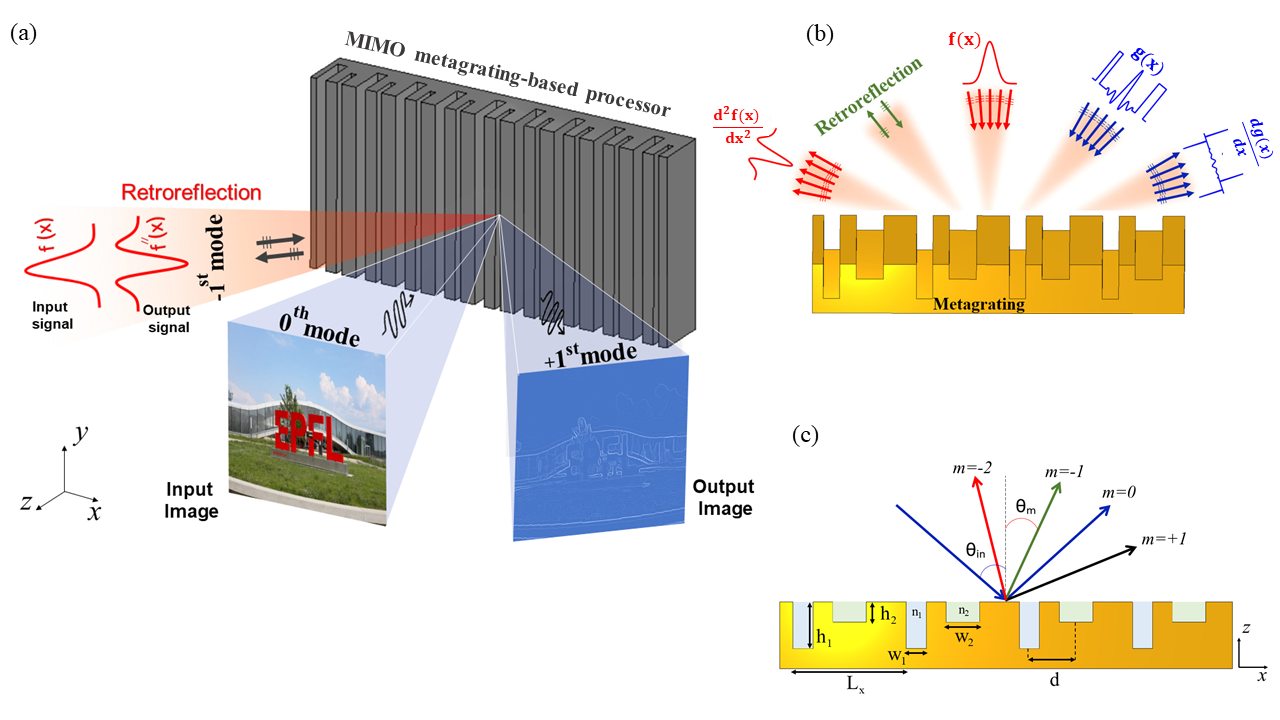}
	\caption{\textbf{Metagrating-based parallel analog signal processing. } (a) Illustration of wave-based  analog signal processing and edge detection through the high-order Floquet modes provided by an all-metallic metagrating. (b) Different channels for parallel analog computing in high-order Floquet harmonics. (c) 2D view of the compound metallic grating with two grooves per period and the associated high-order Floquet harmonics.}\label{fig:epsart}
\end{figure}
In this study, we reveal that a suitably engineered all-metallic metagrating can open multiple non-specular channels for parallel implementation of symmetric and asymmetric optical transfer functions at both normal and oblique illuminations. Closed-form expressions are presented to predict the reflection phase and amplitude information of the space channels opened by the designed compound metallic grating. The application of the designed metagrating for accomplishing first- and second-order differentiation as well as detecting sharp edges of an input image is investigated through several illustrative demonstrations. 
\section{Fundamental Theory of Compound Metallic Grating}
The main idea of this paper is graphically depicted in \textcolor{blue}{Fig. 1a}, where a single metagrating is responsible for realizing  parallel mathematical operations on the input signals coming from different directions. The symmetric geometry originated from the presence of a single groove in each period does not allow to realize the asymmetry required for performing odd-order transfer functions. Thus, we consider a two-dimensional (2D) sparse array (${\partial }/{\partial y}$ $=$ $0$) composed of two grooves in each period of a metallic medium filling the half-space $z$$<$$0$, as shown in \textcolor{blue}{Figs. 1b, c}. The grooves are specified by the widths of $w_1$, $w_2$ and center-to-center distance $d$, heights $h_1$, $h_2$, while being filled with dielectrics of permittivity values $\epsilon_{r1}$ $(=n_1^2)$ and $\epsilon_{r2}$ $(=n_2^2)$, respectively. The whole structure is surrounded by a medium with dielectric constant $\varepsilon_{r0}$ $(=n_0^2)$. The position and dimensions of the grooves can be set at will and the periodicity along $x$ axis is indicated by $L_x$. The metagrating is illuminated by a general oblique 2D beam profile with the angle $\theta_{in}$ and the transverse-magnetic (TM) polarization. By leveraging  the superposition principle and spectral decomposition, the input $f_{in}$ and output $f_{out}$ optical signals can be expanded based on an infinite set of plane waves, all traveling in different directions with wavenumbers $k_x$: 
\begin{subequations}
		\begin{eqnarray}
f_{in}(x,z)=\int\limits_{k_{0x}-W}^{k_{0x}+W}{\tilde{F}_{\text{inc}}^{y}\left( {{k}_{x}}\right)} ~\text{exp}\,(-j{{k}_{x}}{{x}_{\text{}}}+j{{k}_{z}}{{z}_{\text{}}})\,d{{k}_{x\text{}}} \\
f_{out}(x,z)=\int\limits_{k_{0x}-W}^{k_{0x}+W}\tilde{F}_{\text{inc}}^{y}\left( {{k}_{x}}\right){\tilde{H}_{\text{ref}}^{y}\left( {{k}_{x}}\right)} ~\text{exp}\,(-j{{k}_{x}}{{x}_{\text{}}}-j{{k}_{z}}{{z}_{\text{}}})\,d{{k}_{x\text{}}}
  \end{eqnarray}
\end{subequations}

Here, $W$ denotes the beamwidth of the input field and the harmonic time dependency $e$$^{j \omega t}$ is omitted. The spatial frequency content of the incident field is represented by the various plane wave amplitudes  around $k_x$=$k_0\sin(\theta)$, that together form the input signal. The compound metallic grating, depicted in \textcolor{blue}{Fig. 1c}, interacts differently with each of these plane wave components. A Rayleigh expansion can be performed to write the total fields at the upper half-space: 
\begin{subequations}
		\begin{eqnarray}
\tilde{H}_{{z>0}}^{y}= {e^{j{k_{z0}}z}}{e^{ - j{k_{x10}}x}} + \sum\limits_m {\tilde{R}_m {e^{ - j{k_{zm}}z}}{e^{ - j{k_{xm}}x}}} \\
\tilde{E}_{{z>0}}^{x}=  - {Y _{0}}{e^{j{k_{z0}}z}}{e^{ - j{k_{x0}}x}} + \sum\limits_m {{Y _{m}}\tilde{R}_m{e^{ - j{k_{zm}}z}}{e^{ - j{k_{xm}}x}}}
		\end{eqnarray}
\end{subequations}

\begin{subequations}
in which, the subscript $m$ corresponds to the order of space harmonics, and
\begin{eqnarray}
& k_{xm}=k_{x0}+2m\pi/L_x \\ 
& k_{zm}=-jk_{0} \sqrt{(n_0 sin(\theta_i)+m\lambda/L_x)^2-n_0^2} 
\end{eqnarray}
\end{subequations}

Here, $k_{x0}=k_0 n_0 sin(\theta_i)$ and $ Y_{m}={k_{zm}} / {\omega \varepsilon_0 n^2_0}$ denote respectively the x-directed wavenumber the admittance of the $m^{th}$ TM-polarized diffractive mode in the upper half-space. After applying the proper boundary conditions and solving the related equations, the specular and non-specular reflection coefficients can be expressed as: 
\begin{subequations}
	\label{Reflection coefficients}
	\begin{equation}
	\label{R_0}
	\tilde{R}_0=\frac{2}{Y_{0}}\frac{A_0}{B}+1
	\end{equation}
	\begin{equation}
	\label{R_n}
	\tilde{R}_{m\ne0}=\frac{2}{Y_{m}}\frac{A_m}{B}
	\end{equation}
\end{subequations}

in which, $A$, $A_m$, and $B$ are complex numbers. A detailed derivation of the mathematical expressions of these coefficients can be found in \textcolor{blue}{Supplementary Information A}. It should be noted that \textcolor{blue}{Eqs. (4a), (4b)} include both phase and amplitude information of the diffractive modes when the metagrating is excited by an oblique plane wave with an arbitrary angle of incidence.

\section{Results and Discussion}

\subsection{Unlocked Channels}
As shown by \textcolor{blue}{Eq. (3a)}, the scattered wave is a discrete superposition of space harmonics, which can be propagative or not, depending on the metagrating design. Upon illuminating by an oblique plane wave with $\theta_{in}$, the reflected wavefronts are oriented along $\theta_{out}$ in such a way that \cite{behroozinia2020real,rajabalipanah2021analytical}: 
\begin{align}
& \sin \theta_{out,m}=\sin \theta_{in}+m\lambda/L_x  ~~~(m=0, \pm1, \pm2, ...)
\end{align}

In a multi-channel configuration like \textcolor{blue}{Fig. 1b}, those channels can serve to perform optical analog computing. We should remark that in our study, the metagrating is excited by oblique beam profiles of beamwidth $W$, and thus, \textcolor{blue}{Eq. (5)} must be evaluated for all incident wave angles within $\theta_{in}\pm\arcsin(W/k_0)$. With simple algebraic manipulations on \textcolor{blue}{Eq. (5)}, the unlocking condition for each diffractive mode turns into: 
\begin{align}
& \left| \sin {{\theta }_{in}}\sqrt{1-{{\xi }^{{}}}}\pm \cos {{\theta }_{in}}\xi +\frac{m}{{{L}_{nx}}} \right|<1
\end{align}

\begin{figure}[h]
	\centering
	\includegraphics[height=4.4in
	]{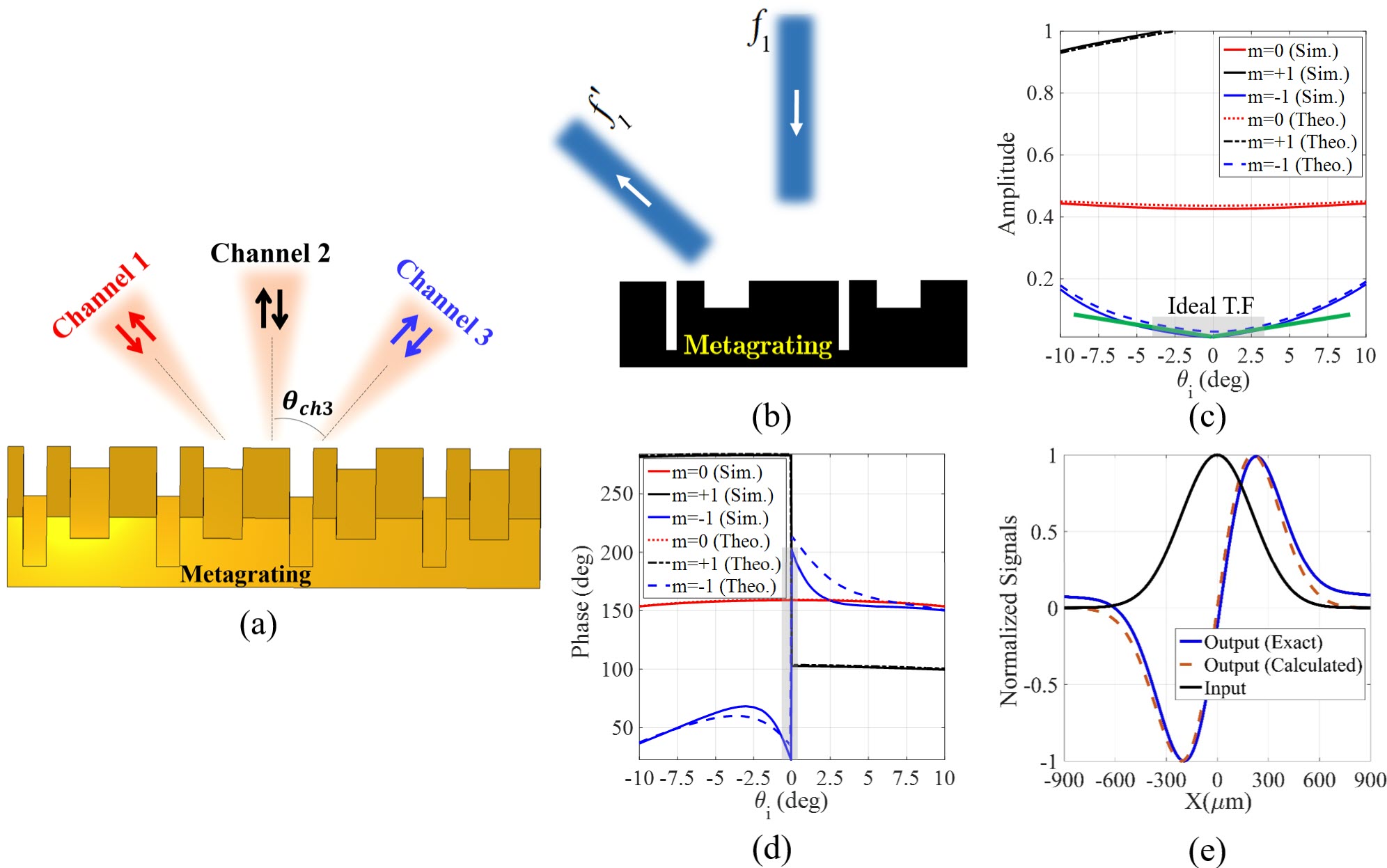}
	\caption{\textbf{Single-operator Metagrating processor. } (a) Shematic illustration of the three processing channels and (b) a single-operator metagrating for performing first-order spatial differentiation in a non-specular reflection mode. The optimum structural parameters are $\varepsilon_{r1}$ $=\varepsilon_{r2} $ $=\varepsilon_{r0} =1$, $w_1=0.288L_x$, $w_2=0.145L_x$, $h_1=0.17L_x$, $h_2=0.087L_x$, and $d=0.257L_x$. The angular spectra of the (c) amplitude and (d) phase for different spatial harmonics. (e) The input field and the corresponding exact/calculated output signal.   }\label{fig:epsart} 
\end{figure}

wherein, $\xi=W/k_0$ and $L_{nx}=L/\lambda$. Given the purpose of design and according to \textcolor{blue}{Fig. 2a}, working with three accessible channels: $m=\{-1,0,1\}$ for $\theta_{in}=0$, $m=\{0,-1, -2\}$ channels for $\theta_{in}$$<$$0$, and $m=\{0,1, 2\}$ channels for $\theta_{in}$$>$$0$ is sufficient and the higher-order modes must be kept evanescent in each case. Taking $\xi=0.2$, \textcolor{blue}{Supplementary Figs. S1a-f} show the solution domain for different FB modes based on \textcolor{blue}{Eq. (6)}, disclosing the best choice for grating periodicity corresponding to the desired angle of incidence. As designed, only three FB modes $m=0,\pm1$ will be propagating for the input beam profile, and the rest will be evanescent, as long as the periodicity and the incident wave angle satisfy $\lambda<L_x<1.5\lambda$ and $-50^\circ<\theta_{in}<50^\circ$, respectively. 

\begin{figure*}[h]
	\centering
	\includegraphics[width=\textwidth]{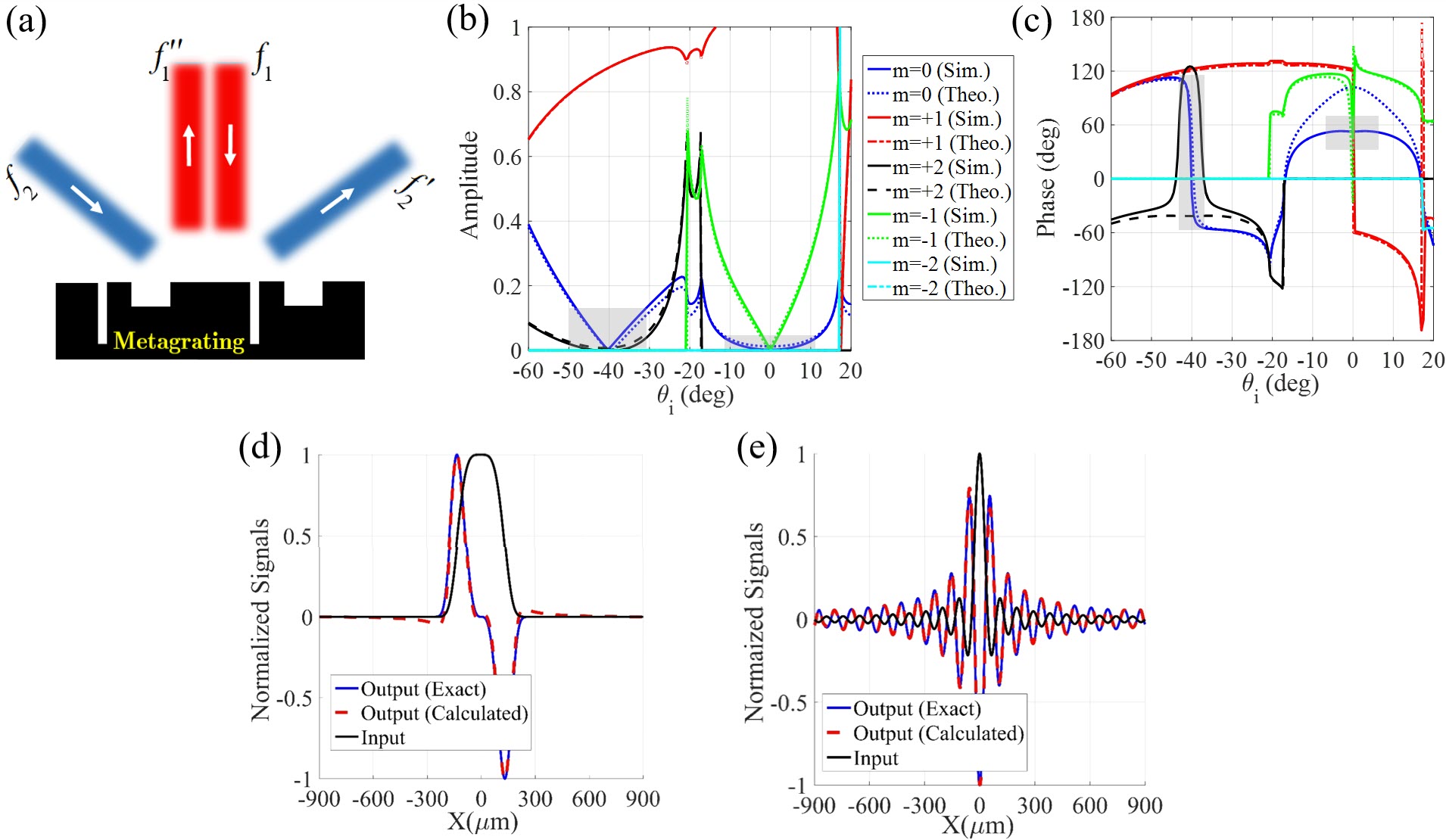}
	\caption{\textbf{Dual-operator Metagrating processor. }  (a) The schematic illustration of a dual-operator metagrating for performing first- and second-order spatial differentiation in specular reflection modes. The optimum structural parameters are $\varepsilon_{r1}$ $=\varepsilon_{r2} $ $=\varepsilon_{r0} =1$, $w_1=0.035L_x$, $w_2=0.0385L_x$, $h_1=0.467L_x$, $h_2=0.45L_x$, and $d=0.643L_x$. The angular spectra of the (b) amplitude and (c) phase for different spatial harmonics. (d), (e) The input fields and the corresponding exact/calculated output signals.   }\label{fig:epsart}
\end{figure*}
Hereafter, we intend to show  how the multiple channels provided by the  metagrating can be exploited for performing analogue signal processing and be used for realizing different functionalities at the same time.

\begin{figure*}[t]
	\centering
	\includegraphics[width=\textwidth]{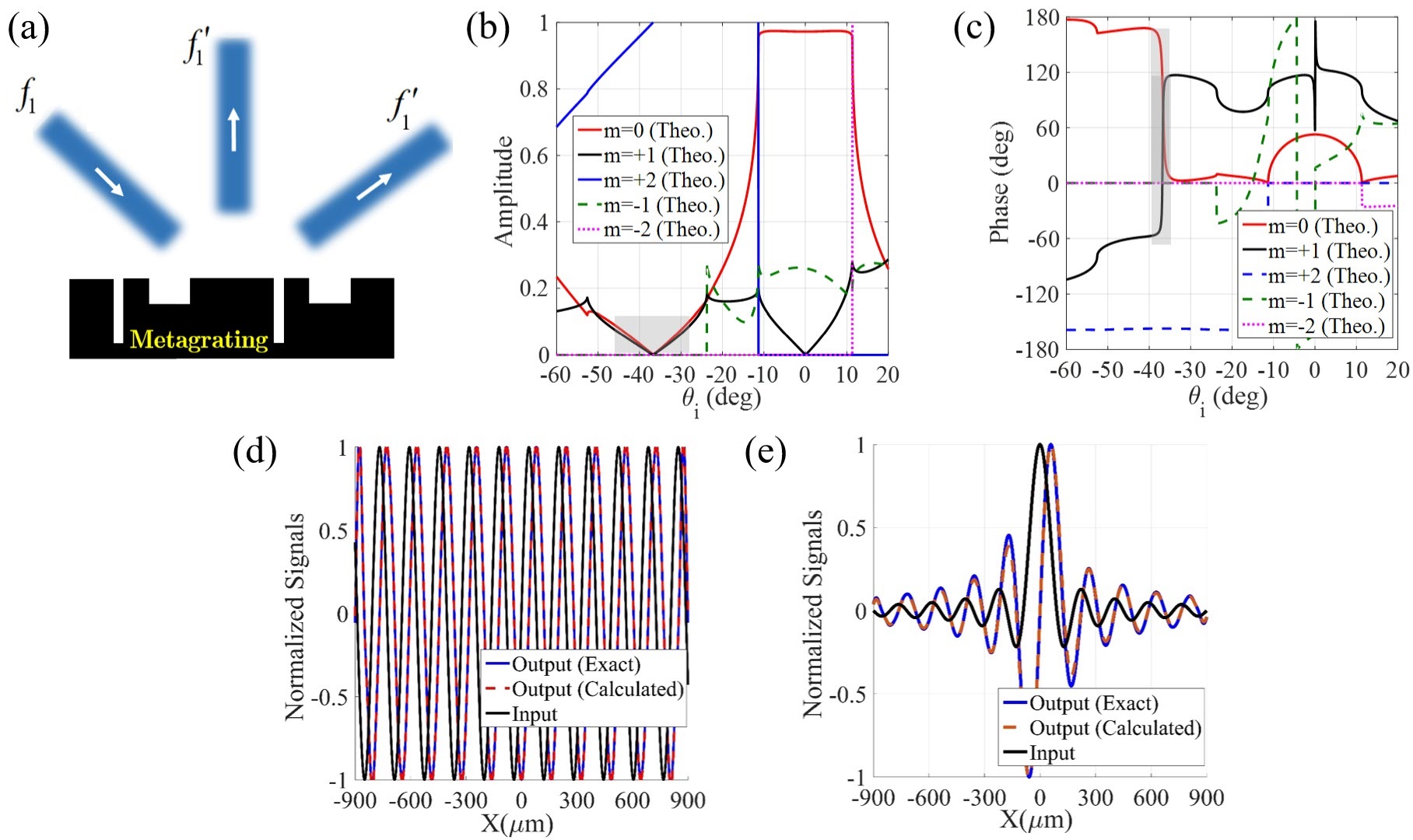}
	\caption{\textbf{Single-input Multi-output Metagrating processor. } (a) The schematic illustration of a single-input dual-output metagrating for performing first-order spatial differentiation in both specular and non-specular reflection modes. The optimum structural parameters are $\varepsilon_{r1}$ $=\varepsilon_{r2} $ $=\varepsilon_{r0} =1$, $w_1=0.026L_x$, $w_2=0.0476L_x$, $h_1=0.43L_x$, $h_2=0.427L_x$, and $d=0.538L_x$. The angular spectra of the (b) amplitude and (c) phase for different spatial harmonics. (d), (e) The input fields and the corresponding exact/calculated output signals.   }\label{fig:epsart}
\end{figure*}

\subsection{Illustrative Examples}

\textbf{Single-Operator Metasurface}. We adjust the periodicity of the metagrating so as to provide three active channels (see \textcolor{blue}{Fig. 2a}).  Accordingly, we set the periodicity of the designed metagrating as $L_x$=1.3$\lambda_0$ ($f_0$=1 THz), which orients the channels along $\theta$=$\pm40^\circ$ directions at $f$=1.2 THz. Once the input and output channels are determined, the width and height of the contributing grooves are optimized so that the angular dispersion of the scattering coefficient between these channels in \textcolor{blue}{Eqs. (10a), (10b)}, emulates the $k_x$-dependency of desired transfer functions. For instance, we can achieve 1$^{\text{st}}$- and 2$^{\text{nd}}$-order spatial differentiation of the input signal exciting the metasurface from the $i^{th}$ channel and exiting at the $j^{\text{th}}$ channel, provided that  $S_{ji}(\theta)$=$jk_0(\sin(\theta)-\sin(\theta_{\text{inc}}))$ and $S_{ji}(\theta)$=$-k_0^2(\sin(\theta)-\sin(\theta_{\text{inc}}))^2$, respectively. Here, our goal is to design a periodic surface that applies 1$^{\text{st}}$-order spatial differentiation operation on the input signal traveling from port 2 to port 1 (see \textcolor{blue}{Fig. 2b}). A comprehensive parametric study based on the theoretical representation of \textcolor{blue}{Eqs. (4a), (4b)}, has been carried out to find the optimum parameters of the metagrating and minimize the following error function: 
\begin{equation}
\label{PEC_bc_2}
E=\sum\limits_{q}^{{}}{\left[ {{\left( \left| {{S}_{ij}}\left( {{\theta }_{q}} \right) \right|-\left| \tilde{H}\left( {{\theta }_{q}} \right) \right| \right)}^{2}}+{{\left( \angle {{S}_{ij}}\left( {{\theta }_{q}} \right)-\angle \tilde{H}\left( {{\theta }_{q}} \right) \right)}^{2}} \right]}
\end{equation}

\begin{figure*}[h]
	\centering
	\includegraphics[width=\textwidth]{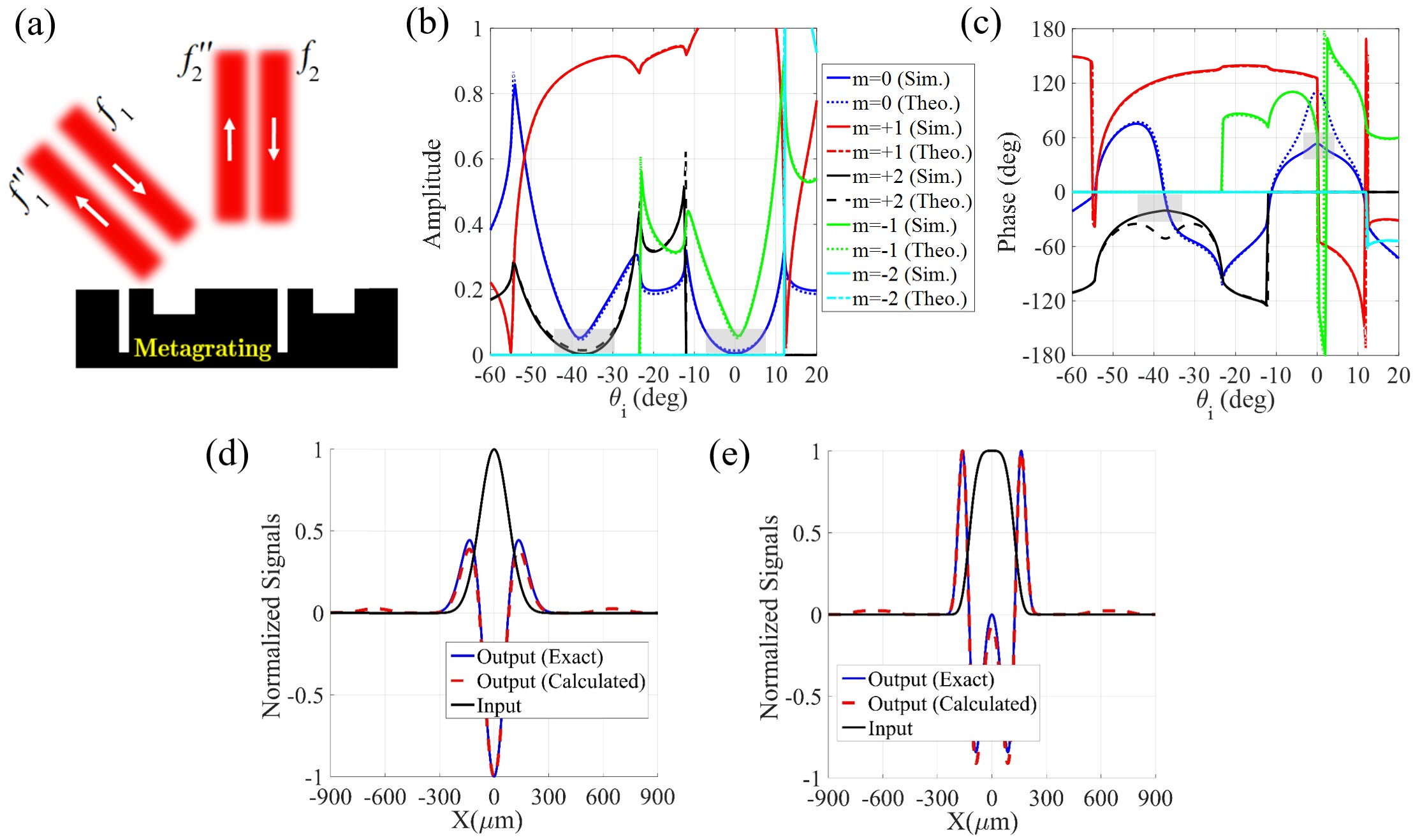}
	\caption{\textbf{Retro-directive Processing channels. } (a) The schematic illustration of a dual-channel metagrating for performing second-order spatial differentiation in both specular and non-specular (retro-reflection) reflection modes. The optimum structural parameters are $\varepsilon_{r1}$ $=\varepsilon_{r2} $ $=\varepsilon_{r0} =1$, $w_1=0.0863L_x$, $w_2=0.069L_x$, $h_1=0.7336L_x$, $h_2=0.71L_x$, and $d=0.6381L_x$. The angular spectra of the (b) amplitude and (c) phase for different spatial harmonics. (d), (e) The input fields and the corresponding exact/calculated output signals.  }\label{fig:epsart}
\end{figure*}

Here, $\tilde{H}(\theta)$ indicates the transfer function of choice, $i$ and $j$ refer to the output and input ports, respectively, and the above summation is calculated over $q$ discrete angles in the vicinity of the incident wave angle. The angular spectra of $|S_{12}(\theta)|$ and $\angle S_{12}(\theta)$ are plotted in \textcolor{blue}{Figs. 2c, d}, respectively, and the optimum geometrical parameters are given in the caption of the same figure. From this figure, one can immediately deduce that the non-specular $S_{12}$ channel ($m=-$1) admits a linear trend and an asymmetric $180^\circ$ phase jump around $\theta=0$, which define the transfer function of $1^{\text{st}}$-order differentiation. To evaluate the performance of our single-operator metasurface differentiator, the Gaussian signal of \textcolor{blue}{Fig. 2e} is utilized to launch the metagrating from port 2 and the calculated output field is shown in the same figure. The result is compared with the exact response indicating that the output field at port 1 is indeed the response expected from a first-order derivative operation, with only a 2$\%$ error, as defined by \textcolor{blue}{Eq. (7)}. Indeed, the designed metagrating successfully deflects the first-order derivative of the normally incident signals into $\theta$=$-40^\circ$ direction without using any additional beam splitting sub-block. Conversely, previous metasurfaces based on the GF method cannot provide a reciprocal solution for asymmetric on-axis transfer functions at the reflection side \cite{momeni2020reciprocal}. Our structure therefore relaxes vexing complexities of previous designs that needed oblique illumination setups and beam splitting devices \cite{momeni2020reciprocal}.

\textbf{Multi-Operator Metasurface}. Up to now, the proposed metagratings realize only a single processing processing operation on a single channel. Hereafter, we intend to design metagratings which successfully create multiple channels, each of which enables different processing or scattering functionalities. To this aim, the optimization procedure of \textcolor{blue}{Eq. (7)} is simultaneously accomplished for different scattering parameters, \textit{e.g.,} $S_{ij}$ and $S_{uv}$. To exemplify how the operation frequency can be adjusted, we decided to change the periodicity of the structure to $L_x=1.22 \lambda_0$, so that the orientation of $m=\pm1$ channels points in the $\theta=40^\circ$ direction for normal excitations at a higher frequency, namely $f$=1.3 THz. Both specular and non-specular channels can be involved and we carry out the numerical optimizations for different computing scenarios displayed in \textcolor{blue}{Figs. 3a, 4a, 5a}. In the first demonstration, the metagrating is parametrically adjusted so that the $S_{13}(\theta)$ (specular) and $S_{22}(\theta)$ (specular) coefficients implement first- and second-order differentiation operations, respectively (see \textcolor{blue}{Fig. 3a}). The synthesized transfer functions are plotted in \textcolor{blue}{Figs. 3b, c}. A parabolic shape is acquired for $|S_{22}(\theta)|$ around $\theta$=$0^\circ$ while no phase change is noticed. At the same time, $S_{13}(\theta)$ provides an amplitude null with 180$^\circ$ phase jump around $\theta$=$-40^\circ$.  
Two input fields, with profiles varying as exp$(-\alpha x^4)$ and sinc$(\beta x)$, are employed to excite the designed metagrating from ports 1 and 2, respectively ($\alpha=0.008$ and $1/\beta=6\times 10^8$ are two constants).
Using \textcolor{blue}{Eqs. (1a), (1b)} and the achieved scattering coefficients, the output fields leaving the metagrating from ports 2 and 3 are calculated and presented in \textcolor{blue}{Figs. 3d, e}, respectively. As seen, the output signals at ports 2 and 3 are nothing but the first- and second-order derivative of the input fields at ports 1 and 2, respectively. The obtained results indicate that a suitably-designed metagrating successfully provides multi-input multi-output channels, each of which, enables different processing functionalities. Our purpose in the second example is to design a single-input multi-output metagrating manifesting the angular dispersion of $jk_0\sin\theta$ (first-order spatial differentiation) in its $S_{21}(\theta)$ (non-specular) and $S_{31}(\theta)$ (specular) scattering coefficients (see \textcolor{blue}{Fig. 4a}). The transfer functions realized by the optimized metagrating are shown in \textcolor{blue}{Figs. 4b, c} from which, one can immediately deduce that both $S_{21}(\theta)$ and $S_{31}(\theta)$ parameters expose a linear trend and an asymmetric $180^\circ$ phase jump near their corresponding angles, consistent with the transfer function of the $1^{\text{st}}$-order differentiation operation. The structure has one input field but the output fields at ports 2 and 3 are examined by  input signals with profiles of sin($\alpha x$) and sinc($\beta x$), respectively. The input and output fields displayed in \textcolor{blue}{Figs. 4d, e} verify that this single-input multi-output processing mission is perfectly accomplished by the designed metagrating. For a quantitative comparison, the exact output responses are also illustrated. As a final demonstration example, we form a multi-operator metagrating performing second-order spatial differentiation for the input signals coming from either port 1 or 2 (see \textcolor{blue}{Fig. 5a}). \textcolor{blue}{Fig. 5b, c} demonstrate the angle-dependent amplitude and phase of the transfer function of the designed metagrating. All indicator features of the second-order differentiation operation are observable in $S_{22}(\theta)$ (specular) and $S_{11}(\theta)$ (non-specular) around $\theta$=0$^\circ$ and $\theta$=40$^\circ$, respectively.  \textcolor{blue}{Figs. 5d, e} depicts the input fields, exp$(-\alpha x^2)$ and exp$(-\alpha x^4)$, and the corresponding output signals, confirming that the designed multi-operator metagrating is capable of independently processing two input signals. 

\begin{figure*}[h]
	\centering
	\includegraphics[width=\textwidth]{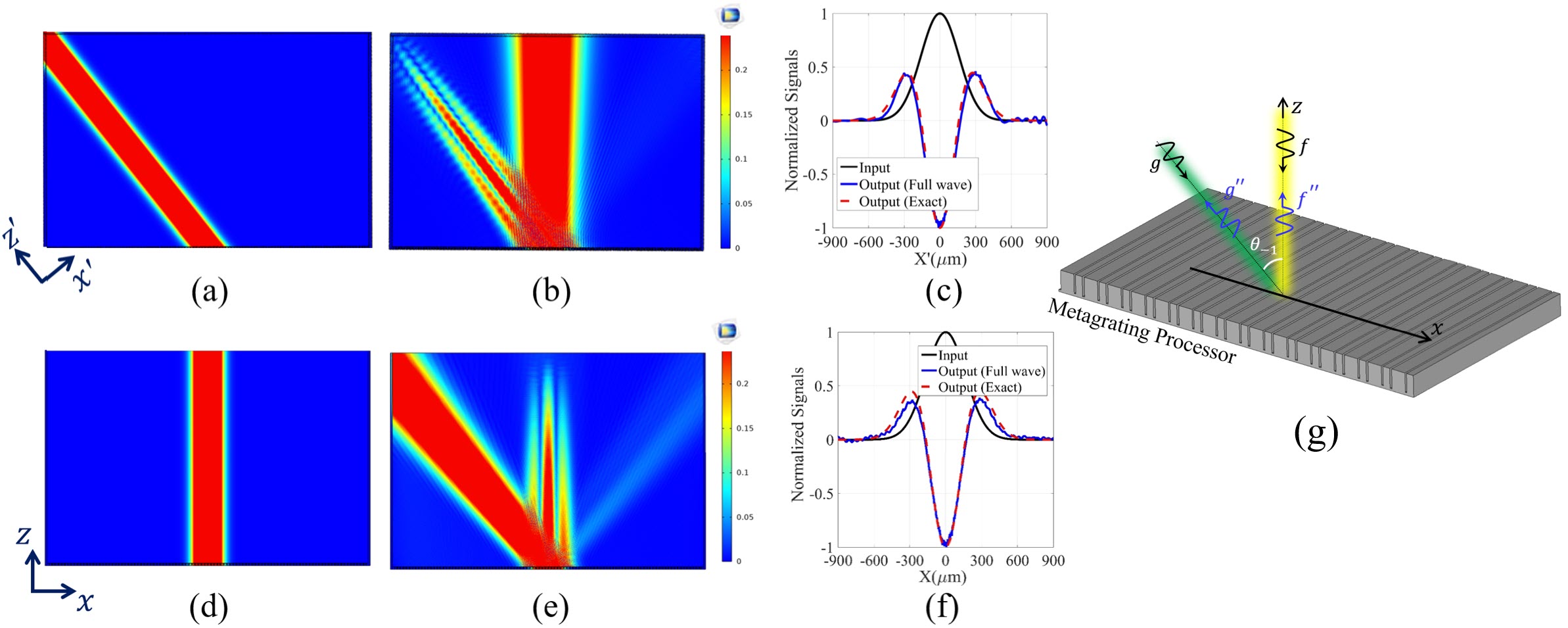}
	\caption{\textbf{Full-wave verification  of non-specular processing channels.} Full-wave verification of the proposed signal processing idea for the designed metagrating of \textcolor{blue}{Fig. 6a}. (a), (d) The obliquely- and normally-oriented input Gaussian-shape beams, (b), (e)  the corresponding reflected fields, and (c), (f) cut-line of the reflected fields along with the exact second-order derivative of the input fields. (g) The 3D schematic view of the studied metagrating. }\label{fig:epsart}
\end{figure*}

To further validate the performance of the proposed metagratings, full-wave simulations have been carried out through COMSOL Multiphysics. The configuration is shown in \textcolor{blue}{Fig. 6g} in which two TM-polarized Gaussian-shape beams illuminate the metagrating of \textcolor{blue}{Fig. 6a} from two different directions $\theta=0^\circ$ and $\theta=40^\circ$ (see \textcolor{blue}{Figs. 6a, d}). The medium surrounding the designed metagratings is filled by air and the boundary conditions are selected as perfect match layer (PML). \textcolor{blue}{Figs. 6b, d} demonstrate the scattered fields for each of normal and oblique illuminations. A cut-line of the scattered fields in each case is plotted in \textcolor{blue}{Figs. 6c, f} indicating that the output signals successfully obey the exact version of the second-order spatial derivative of the Gaussian-shape beams. Thus, the full-wave simulations verify well our idea and analytical model, confirming  the relevance of high-order Floquet modes to perform analog signal processing.   

Finally, a concrete 1D edge detection application is evaluated by using the metagrating differentiator of \textcolor{blue}{Fig. 2b}, where "Advanced Optical Material" logo is utilized as input image (see \textcolor{blue}{Fig. 7a}). The reflected image displayed in \textcolor{blue}{Fig. 7b} highlights all edges of the incident image along the vertical direction. Therefore, the proposed multi-operator metagrating can also be an important key in parallel image processing applications, including in THz imaging.

\begin{figure}[h]
	\centering
	\includegraphics[width=8cm]{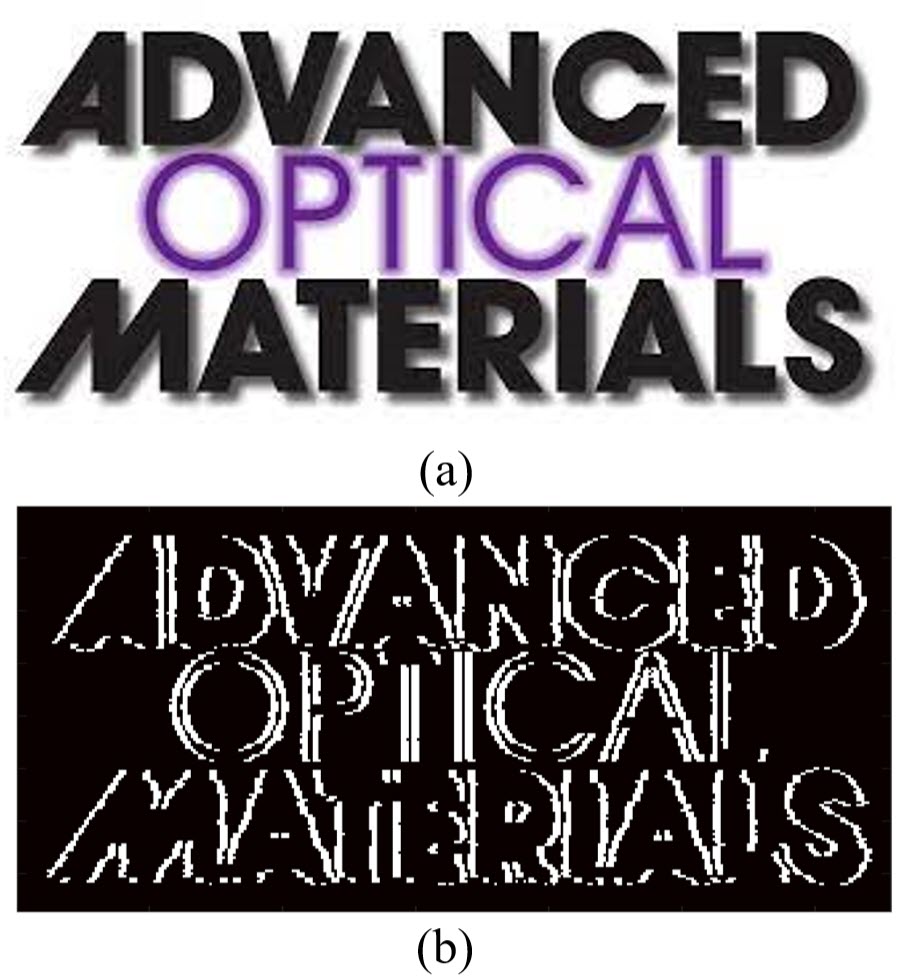}
	\caption{\textbf{Edge detection demonstration of the metagrating processor}. Illustration of the edge detection application by using the metagrating of \textcolor{blue}{Fig. 2b}. (a) "Advanced Optical Material" logo as the input image, (b) the reflected image captured at the first-order space harmonic, demonstrating the vertical edges of the input image. }\label{fig:epsart}
\end{figure}

\section{Conclusion}
~~In summary, we exploited high-order spatial harmonics of a multi-functional angle-multiplexed metagrating to realize various scattering and signal processing functionalities at the same time. Analytical expressions were presented for engineering the phase and amplitude information of each spatial harmonic. We demonstrated metagrating configurations that implement diverse optical computing functions such as first- and second-order differentiation operations, and several scattering functionalities such as retro-reflection when excited from different channels. Full-wave finite-element simulations confirmed our theory. Using the proposed angle-multiplexed metagratings, we completely escape the problem of realizing asymmetric optical transfer functions for normal illuminations, avoiding bulky additional splitting blocks to separate the routes of the incident and reflected signals, and allowing for much more compact configurations compared with previous attempts. Besides, the sparsity of our wavelength-scale array significantly relaxes fabrication tolerance, which is a crucial advantage to move the concept towards optical frequencies. This new class of multi-functional processors may find great potential applications in integrated photonic devices and imaging systems that process optical signals coming from different directions at the speed of light, as they reflect onto structured surfaces. 


\textbf{Acknowledgements}:
A. Momeni and R. Fleury acknowledge funding from the Swiss National Science Foundation under the Eccellenza grant number 181232.



\bibliography{cas-refs}

\newpage 
\section{SUPPLEMENTARY MATERIAL}

Each groove can be modeled by a parallel plate waveguide supporting a single transverse electromagnetic (TEM) wave below the cutoff frequency $f_c$$=$$c/ 2\max[w_1 n_1, w_2n_2]$. Subsequently, we derive the tangential electric fields according to Maxwells equations as a superposition of forward and backward propagating plane waves inside each of grooves, and use mode matching on the aperture $z=0$ to retrieve the scattering coefficients and vanishing of the tangential electric field at the PEC interface:

\begin{subequations}
	\label{RegionII}
	\begin{equation}
	\label{HfieldRegionII}
	{H_{\text{grv1}}^{y}} = \sum H_{0,\text{grv1}}^ \pm {e^{\pm j{\beta _{\text{grv1}}}z}}
	\end{equation}
	\begin{equation}
	\label{EfieldRegionII}
	{E_{\text{grv1}}^{x}} =  \sum (-1)^{\pm} {Y _{0,\text{grv1}}}H_{0,\text{grv1}}^ \pm {e^{\pm j{\beta_{\text{grv1}}}z}} 
	\end{equation}
\end{subequations}
for $x\in \text{grv1}$, and
\begin{subequations}
	\label{RegionII}
	\begin{equation}
	\label{HfieldRegionII}
	{H_{\text{grv2}}^{y}} = \sum H_{0,\text{grv2}}^ \pm {e^{\pm j{\beta _{\text{grv2}}}z}}
	\end{equation}
	\begin{equation}
	\label{EfieldRegionII}
	{E_{\text{grv2}}^{x}} =  \sum (-1)^{\pm} {Y _{0,\text{grv2}}}H_{0,\text{grv2}}^ \pm {e^{\pm j{\beta_{\text{grv2}}}z}} 
	\end{equation}
\end{subequations}
for $x\in \text{grv2}$, where ${\beta}_{\text{grvi}}=k_0n_i\sqrt{1+{(1-j)\delta_s}/{w_i}}$ and $ Y_{0,\text{grv,i}}={\beta_{\text{grvi}}} / {\omega \varepsilon_0 n^2_i}$ $(i=1,2)$ are the propagation constant and the admittance of the TEM mode involved inside the first and second grooves. The propagation constant incorporates the ohmic losses of the lossy parallel plate waveguide by making a good conductor approximation with a strong skin effect condition $\delta_s=\sqrt{2/\omega \mu_0 \sigma}$. The continuity of total tangential electric field on the whole surface of each period yields: 


\begin{subequations}
	\label{BCforE}
	\begin{eqnarray}
	\label{BCforE_0 order}
1  - \tilde{R}_0  = \frac{{{Y _{0,\text{grv2}}}}}{{{Y _{0}}}}H_{0,\text{grv2}}^ + M_{\text{grv2} + }^0 - \frac{{{Y _{0,\text{grv2}}}}}{{{Y _{0}}}}H_{0,\text{grv2}}^ - \,M_{\text{grv2} + }^0  \nonumber\\ + \frac{{{Y _{0,\text{grv1}}}}}{{{Y _{0}}}}H_{0,\text{grv1}}^ + M_{\text{grv1} + }^0\, - \frac{{{Y _{0,\text{grv1}}}}}{{{Y _{0}}}}H_{0,\text{grv1}}^ - \,M_{\text{grv1} + }^0
	\end{eqnarray} 
	\begin{eqnarray}
	\label{BCforE_m order}
	\tilde{R}_{m\ne0} =  - \frac{{{Y _{0,\text{grv1}}}}}{{{Y _{m}}}}H_{0,\text{grv1}}^ + M_{\text{grv1} + }^m + \frac{{{Y _{0,\text{grv1}}}}}{{{Y _{m}}}}H_{0,\text{grv1}}^ - M_{\text{grv1}+}^m \nonumber \\ - \frac{{{Y _{0,\text{grv2}}}}}{{{Y _{m}}}}H_{0,\text{grv2}}^ + M_{\text{grv2}+}^m + \frac{{{Y _{0,\text{grv2}}}}}{{{Y _{m}}}}H_{0,\text{grv2}}^ - M_{\text{grv2}+}^m ~ ~ ~ m \ne 0
	\end{eqnarray}
\end{subequations}

where,
\begin{subequations}
	\label{M+-}
	\begin{equation}
	\label{M2+-}
	M_{\text{grv1} \pm }^m = \,\frac{1}{L_x}\int\limits_\text{grv1}^{{}} {{e^{ \pm j{k_{xm}}x}}dx} \,
	\end{equation}
	\begin{equation}
	\label{M3+1}
	M_{\text{grv2} \pm }^m = \frac{1}{L_x}\int\limits_{\text{grv2}}^{} {{e^{ \pm j{k_{xm}}x}}dx} 
	\end{equation}
\end{subequations} 

Here, we multiply the electric fields by $e^{jk_{xm}x}$ and take the integral of both sides over one period. Using the boundary conditions of the tangential magnetic field and then, taking the integral of both sides over each slit width, we have:  
\begin{subequations}
	\label{BCforH}
	\begin{equation}
	\label{BCforH_2}
	L_xM_{\text{grv1} - }^0 + L_x\sum\limits_m {\tilde{R}_m M_{\text{grv1} - }^m}  = {w_1}H_{0,\text{grv1}}^ +  + {w_1}H_{0,\text{grv1}}^ -
	\end{equation}
	\begin{equation}
	\label{BCforH_3}
	L_xM_{\text{grv2} - }^0 + L_x\sum\limits_m {\tilde{R}_m M_{\text{grv2} - }^m}  = {w_2}H_{0,\text{grv2}}^ +  + {w_2}H_{0,\text{grv2}}^ -
	\end{equation}
\end{subequations}

The PEC termination at the end of each groove commands: 
\begin{subequations}
	\label{PEC_bc}
	\begin{equation}
	\label{PEC_bc_2}
	H_{0,\text{grv1}}^ -  = H_{0,\text{grv1}}^ + {e^{ - 2j{\beta _{\text{grv1}}}{h_1}}}
	\end{equation}
	\begin{equation}
	\label{PEC_bc_3}
	H_{0,\text{grv2}}^ -  = H_{0,\text{grv2}}^ + {e^{ - 2j{\beta _{\text{grv2}}}{h_2}}}
	\end{equation}
\end{subequations}

The detailed expressions for $A$ and $B$ coefficients of \textcolor{blue}{Eqs. (4a), (4b)} can then be written as: 

	\begin{align}
{A_m} = - M_{\text{grv1}-}^0 M_{2+}^m S_2 C_3 Y_{0,\text{grv1}} - M_{\text{grv1}-}^0 M_{2+}^m S_2 S_3 Y_{0,\text{grv1}}Y_{0,\text{grv2}} \sum\limits_q{\frac{M_{3+}^q M_{3-}^q}{Y_{1q}}}  \\ \nonumber - M_{3-}^0 M_{3+}^m S_3 C_2 Y_{0,\text{grv2}} - M_{3-}^0 M_{3+}^m S_2 S_3 Y_{0,\text{grv1}}Y_{0,\text{grv2}} \sum\limits_q {\frac{M_{2+}^q M_{\text{grv1}-}^q}{Y_{1q}}} \\ \nonumber +M_{3-}^0 M_{2+}^m S_2 S_3 Y_{0,\text{grv1}} Y_{0,\text{grv2}} \sum\limits_q{\frac{M_{\text{grv1}-}^q M_{3+}^q}{Y_{1q}}} + M_{\text{grv1}-}^0 M_{3+}^m S_2 S_3 Y_{0,\text{grv1}} Y_{0,\text{grv2}} \sum\limits_q{\frac{M_{2+}^q M_{3-}^q}{Y_{1q}}} \\
	{B} = S_2 C_3 Y_{0,\text{grv1}} \sum\limits_q{\frac{M_{2+}^q M_{\text{grv1}-}^q}{Y_{1q}}} + S_3 C_2 Y_{0,\text{grv2}} \sum\limits_q{\frac{M_{3+}^q M_{3-}^q}{Y_{1q}}} + C_2 C_3 +  \\ \nonumber S_2 S_3 Y_{0,\text{grv1}} Y_{0,\text{grv2}} (\sum\limits_q{\frac{M_{2+}^q M_{\text{grv1}-}^q}{Y_{1q}}} \sum\limits_q{\frac{M_{3+}^q M_{3-}^q}{Y_{1q}}} - \sum\limits_q{\frac{M_{2+}^q M_{3-}^q}{Y_{1q}}} \sum\limits_q{\frac{M_{2-}^q M_{3+}^q}{Y_{1q}}}) \\
	{S_i} = (1 - {e^{ - 2j{\beta _i}{h_i}}})~~~~(i=2,3) \\
	{C_i} = \frac{w_{i}}{L_x}(1 + {e^{ - 2j{\beta _i}{h_i}}}) ~~~~(i=2,3) 
	\end{align}

\begin{figure*}[h]
	\centering
	\includegraphics[width=\textwidth]{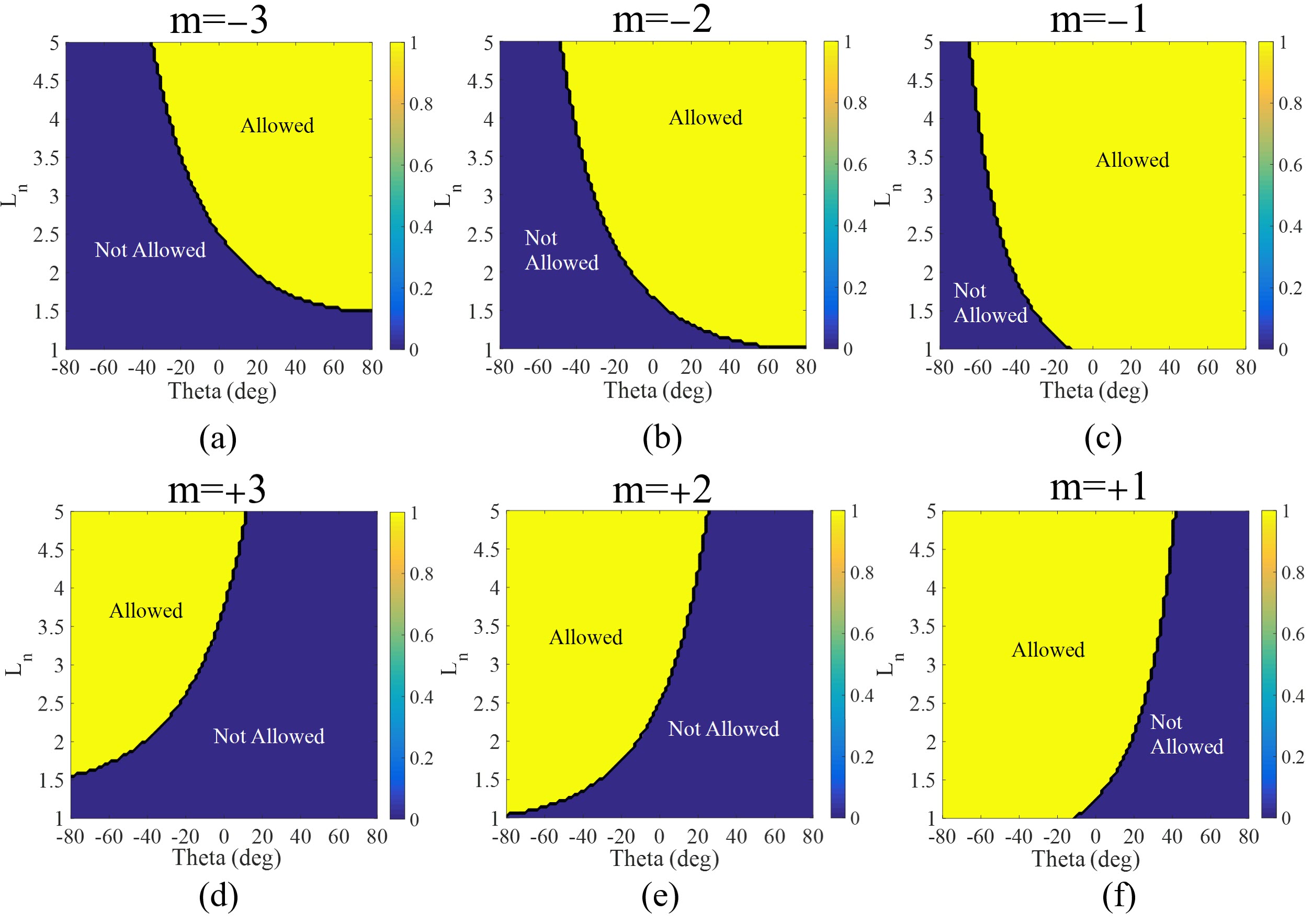}
	\caption{The allowed and not allowed processing channels for different incident wave angles and harmonic numbers where the angular beamwidth of the input beam is specified with $\zeta$=0.2.}\label{fig:epsart}
\end{figure*}

\end{document}